# Computational Imaging Without a Computer: Seeing Through Random Diffusers at the Speed of Light

**Authors:** Yi Luo[1,2,3], Yifan Zhao[1,3], Jingxi Li[1,2,3], Ege Çetintaş[1,2,3], Yair Rivenson[1,2,3], Mona Jarrahi[1,3], Aydogan Ozcan[1,2,3,4,*]

[1] Electrical and Computer Engineering Department, University of California, Los Angeles; Los Angeles, CA, 90095, USA

[2] Bioengineering Department, University of California, Los Angeles; Los Angeles, CA, 90095, USA

[3] California NanoSystems Institute, University of California, Los Angeles; Los Angeles, CA, 90095, USA

[4] Department of Surgery, David Geffen School of Medicine, University of California, Los Angeles; Los Angeles, CA, 90095, USA.

* Corresponding author. Email: ozcan@ucla.edu

## Abstract

Imaging through diffusers presents a challenging problem with various digital image reconstruction solutions demonstrated to date using computers. We present a computer-free, all-optical image reconstruction method to see through random diffusers at the speed of light. Using deep learning, a set of diffractive surfaces are designed/trained to all-optically reconstruct images of objects that are covered by random phase diffusers. We experimentally demonstrated this concept using coherent THz illumination and all-optically reconstructed objects distorted by unknown, random diffusers, never used during training. Unlike digital methods, all-optical diffractive reconstructions do not require power except for the illumination light. This diffractive solution to see through diffusers can be extended to other wavelengths, and might fuel various applications in biomedical imaging, astronomy, atmospheric sciences, oceanography, security, robotics, among others.

**Main Text**

Imaging through scattering and diffusive media has been an important problem for many decades, with numerous solutions reported so far(*1–13*). In various fields, including e.g., biomedical optics(*4, 14*), atmospheric physics(*15, 16*), remote sensing(*17, 18*), astronomy(*19, 20*), oceanography(*21, 22*) as well as autonomous systems and robotics(*23, 24*), the capability to rapidly see through diffusive and scattering media is of utmost importance. In principle, with *a prior* information of the transmission matrix of a diffuser(*5, 25*), the distorted images can be recovered using a computer. However, there is no simple solution to accurately obtain the transmission matrix of a diffuser(*26*). Furthermore, the transmission matrix will significantly deviate from its measured function if there are changes in the scattering medium(*27*), partially limiting the utility of such measurements to see through unknown, new diffusers. To overcome some of these challenges, adaptive optics-based methods have been applied in different scenarios(*4, 9, 28*). With significant advances in wave-front shaping(*29–32*), wide-field real-time imaging through turbid media became possible(*7, 33*). These algorithmic methods are implemented digitally using a computer and require guide-stars or known reference objects, which introduce additional complexity to an imaging system. Digital deconvolution using the memory effect(*34, 35*) with iterative algorithms is another important avenue toward image reconstruction using a computer(*8, 36–39*).

Some of the more recent work on imaging through diffusers has also focused on using deep learning methods to digitally recover the images of unknown objects (*12, 11, 40, 41*). Deep learning has been re-defining the state-of-the-art across many areas in optics, including optical microscopy(*42–47*), holography(*48–53*), inverse design of optical devices(*54–59*), optical computation and statistical inference(*60–69*), among others(*70–72*). To incorporate deep learning to digitally reconstruct distorted images, neural networks were trained using image pairs composed of diffuser-distorted patterns of objects and their corresponding distortion-free images (target, ground truth). Harvesting the generalization capability of deep neural networks, one can digitally recover an image that was distorted by a new diffuser (never seen in the training phase), by passing the acquired distorted image through a trained neural network using a computer(*12*).

In this paper, we present computer-free and all-optical reconstruction of object images distorted by unknown, randomly-generated phase diffusers, as shown in Fig. 1(a). Unlike previous digital approaches that utilized computers to reconstruct an image of the input object behind a diffuser, here we trained a set of diffractive surfaces/layers using deep learning to all-optically reconstruct the image of an unknown object as the diffuser-distorted input optical field diffracts through successive trained layers, i.e., the image reconstruction is processed at the speed of light propagation through the diffractive layers. Each diffractive surface that is trained has tens of thousands of diffractive features (termed as neurons), where the individual phase values of these neurons are adjusted in the training phase through error back-propagation, by minimizing a customized loss function between the ground truth image and the diffracted pattern at the output field-of-view. During this training, many different, randomly-selected phase diffusers, all with the same statistical correlation length, are used to help the generalization of the optical network. After this deep learning-based design of these diffractive layers (which is a one-time effort), they are fabricated to form a physical diffractive network that is positioned between an unknown, new diffuser and the output/image plane. As the input light corresponding to a new, unknown object passes through an unknown diffuser, the scattered light is collected by the trained diffractive network to all-optically reconstruct an image of the object at its output field-of-view, without the need for a computer, any digital computation or an external power source (except for the coherent illumination light).

We validated the success of this approach using coherent THz illumination, and fabricated our designed diffractive networks with a 3D-printer to demonstrate their capability to see through randomly-generated unknown phase diffusers that were never used in the training phase. We also observed an improved object reconstruction quality using deeper diffractive networks that have additional trainable layers. This all-optical image reconstruction achieved by passive diffractive layers enables to see objects through unknown random diffusers and presents an extremely low power solution compared with existing deep learning-based or iterative image reconstruction methods implemented using computers, only requiring power for the coherent illumination source. Learning-based diffractive models presented here to see through diffusers can also work at other parts of the electromagnetic spectrum, including the visible and far/mid-infrared wavelengths. We believe that this framework can bring transformative advances in various fields, where imaging through diffusive media is of utmost importance such as e.g., in biomedical imaging, astronomy, autonomous vehicles, robotics and defense/security applications, among many others.

## Results

**Diffractive network design and experimental set-up**

We designed and 3D-fabricated diffractive networks that can all-optically reconstruct object images that are distorted by random phase diffusers under 400 GHz illumination ($\lambda \approx 0.75$ mm). In terms of the optical set-up, custom fabricated phase diffusers (see the Methods section) are individually placed 40 mm (53$\lambda$) away from the input object plane. The successive diffractive layers (designed for all-optical reconstruction of the object field-of-view) are placed 2 mm away from the diffuser, with a layer-to-layer distance of 2 mm. The output image plane is positioned 7 mm (9.3$\lambda$) from the last diffractive layer along the optical axis (Fig. 1(a)). Based on these parameters, characteristic object distortion generated by a randomly selected phase diffuser is reported in Fig. 1(b). First, we simulated the free-space propagation (FSP) of a distorted object (i.e., seen through a diffuser) without the presence of the diffractive layers, and got its intensity distribution at the output plane, which is shown in the second row of Fig 1(b). Imaging of the same object through the same diffuser by an aberration-free lens is also shown in the third row of Fig 1(b). These images clearly show the impact of the diffuser at the output plane (through free-space propagation or an imaging lens), which makes it impossible to recognize the object unless further computation or digital reconstruction is applied. As we report here, jointly-trained passive diffractive surfaces can perform this needed computation *all-optically*, as the scattered light behind an unknown diffuser passes through these layers, forming an image of the object field-of-view at its output plane, as exemplified in the fourth row of Fig. 1(b).

A diffractive network generalizes to see through unknown, new diffusers by training its layers with numerous image pairs: diffuser-distorted speckle patterns of various input objects and the corresponding distortion-free object images (target). To make our all-optical diffractive system capable of reconstructing an unknown object's image that is distorted by *new* diffusers (i.e., never seen during the training phase), we adopted the strategy of using multiple diffusers to train our diffractive surfaces, following the procedure depicted in Fig. 1(a). All of the diffusers that are used in the training and blind testing phases are assumed to have the same correlation length ($L$~10$\lambda$) and are randomly created as thin phase elements (see the Methods section and Fig. 1(a)). At the beginning of each training epoch, a set of *n* different phase diffusers are initialized to be used throughout the whole epoch. In each iteration within a given epoch, we randomly selected a batch of *B* training images from the MNIST dataset(*73*) (containing 50,000 handwritten digits for training and 10,000 for testing) and used them, one by one, through the amplitude channel of the input object plane. During each iteration, a total of $B \times n$ distorted optical fields were processed by the diffractive network and subsequently measured at the output plane. The corresponding loss value, calculated through a training loss function that blends negative Pearson Correlation Coefficient (PCC) (*11*) and photon loss (see the Methods section), was then used to calculate the error gradients for updating the phase modulation values of the neurons on the diffractive layers, marking the end of one iteration. An epoch was finished when all the 50,000 training images within the MNIST dataset were exhausted to train the network. After being trained for 100 epochs, the network has seen features from a total of $N=100n$ unique phase diffusers that are randomly generated. As demonstrated in the following subsections, this strategy enabled the generalization of the diffractive network to see and reconstruct unknown objects through novel/new phase diffusers that were never used in the training phase.

**All-optical, computer-free image reconstruction through diffusers**

To demonstrate the all-optical image reconstruction performance of a diffractive network, we first trained a 4-layered network using *n*=20 diffusers in each epoch (Fig. 1(c)). After being trained for 100 epochs, the resulting network generalized to an imaging system that can see through diffusers at the speed of light, without the need for a computer or a digital reconstruction algorithm. The trained diffractive layers' phase modulation patterns are reported in Supplementary Fig. S1. To shed light on the operation principles of the trained diffractive network, it was initially tested with new hand-written digits (i.e., MNIST test images that were never used in the training phase) distorted by $n=20$ individual diffusers that were used in the last training epoch (we term these as *known* diffusers, meaning they were used in the training). The first three rows in Fig. 2(a) present the successful all-optical reconstruction results corresponding to these new hand-written digits that were distorted by three (K1, K2 and K3) of the last $n=20$ known diffusers. Next we blindly tested the same trained diffractive network with new phase diffusers that were not used during the training. For this, we randomly generated 20 novel/new diffusers and Figure 2(b) shows the all-optical reconstruction results of the same objects (never seen by the network before) distorted by unknown/new diffusers (B1, B2 and B3), which were randomly selected from the 20 new phase diffusers. A comparison between Figures 2(a) and 2(b) reveals the generalization performance of the trained diffractive network to all-optically reconstruct unknown objects through unknown phase diffusers that were never seen before.

In addition to MNIST test images, we further tested the same diffractive network with resolution test targets having different periods (10.8mm and 12mm respectively); see Fig. 2, last 2 rows. These types of resolution test targets, composed of periodic lines, were never seen by the diffractive network before (which was trained with only MNIST data), and their successful reconstruction at the network output plane (Fig. 2) further supports the generalization of the diffractive network's capability to reconstruct any arbitrary object positioned at the input field-of-view, instead of overfitting to a local minimum covering only images from a specific dataset. To quantitatively analyze the generalization performance of trained diffractive networks, in Figures 3(a) and 3(b) we also report the measured periods corresponding to the all-optically reconstructed images of different resolution test targets that were seen through the last $n$ known diffusers (used in the last training epoch) as well as 20 new, randomly generated diffusers (never used during the training). Despite the use of different training strategies (with $n = 1, 10, 15, 20$), the results reported in Fig. 3 reveal that all these trained diffractive network models can resolve and accurately quantify the periods of these resolution test targets seen through known as well as new/novel diffusers.

After these numerical analyses of all-optical image reconstruction under different conditions, next we experimentally verified its performance and fabricated the designed diffractive layers using a 3D printer (Fig. 1(d)); we also fabricated diffusers K1-K3 and B1-B3 as well as 5 test objects (3 hand-written digits and 2 resolution test targets). The test objects were further coated using aluminum foil to provide binary transmittance. For each hand-written digit, a 42×42 mm field-of-view at the output plane was imaged by scanning a 0.5×0.25 mm detector with a step size of 1 mm in each direction (see Fig. 1(e)). The experimental results are shown in Figs. 4(a) and (b), clearly demonstrating the success of the all-optical network's capability to see through unknown diffusers. For comparison, we also report the intensity distribution generated by a lens-based imaging system as well as free-space propagation (without the diffractive layers) of an input object with the presence of the diffuser K1 (see Supplementary Fig. S2); a similar comparison is also provided in Fig. 1(b). These comparisons clearly highlight the success of the all-optical image reconstruction achieved by the diffractive network despite the presence of significant image distortion caused by the unknown diffuser and free-space propagation.

We also imaged resolution test targets using the same experimental setup at the output plane of the diffractive network (see Fig. 4(c)). From the all-optically reconstructed output images of the diffractive network, we measured the periods of the resolution test targets imaged through known and (*new/novel*) diffusers as 10.851± 0.121mm (*11.233±0.531mm*) and 12.269±0.431mm (*12.225±0.245mm*), corresponding to the fabricated resolution test targets with periods of 10.8mm and 12mm, respectively. These experimental results further demonstrate the generalization capability of the trained diffractive network to all-optically reconstruct/image unknown objects through unknown diffusers, which were never used during the training phase; moreover, we should emphasize that this fabricated diffractive network design was only trained with MNIST image data, without seeing grating-like periodic structures.

**Performance of all-optical image reconstruction as a function of the number of independent diffusers used in the training**

An important training parameter to be further examined is the number of diffusers ($n$) used in each epoch of the training. To shed more light on the impact of this parameter, we compared the all-optical reconstruction performance of four different diffractive networks trained with $n=1$, $n=10$, $n=15$ and $n=20$, while keeping all the other parameters the same. To further quantify the image reconstruction performance of these trained diffractive networks, we adopted the Pearson Correlation Coefficient (PCC) (*74*) as a figure of merit, defined as:

$$P = \frac{\sum(O - \bar{O}) \cdot (G - \bar{G})}{\sqrt{\sum(O - \bar{O})^2 \cdot \sum(G - \bar{G})^2}} \qquad (1)$$

where $O$ is the output image of the diffractive network and $G$ is object image to be reconstructed, i.e., the ground truth. Using this metric, we calculated the mean PCC value for the all-optical reconstruction of 10,000 MNIST test objects (never used in the training) distorted by the same diffusers. Stated differently, after being trained for 100 epochs, all the finalized networks ($n = 1$, 10, 15, 20) were compared to each other by calculating the average PCC values over unknown MNIST test objects distorted by each one of the $100n$ known diffusers as well as each one of the 20 new/novel randomly generated diffusers (see Fig. 5). This figure should *not* be confused with learning curves typically used to monitor the ongoing training of a neural network model; in fact, the results in Fig. 5 report the all-optical reconstruction fidelity/quality achieved for unknown test objects *after* the training is complete. From top to bottom, the four panels in Fig. 5(a) present the comparison of the diffractive networks trained with $n=1$, $n=10$, $n=15$ and $n=20$, respectively, while the inserts in last three panels show the same plot zoomed into the last 50 diffusers. An increased PCC value can be clearly observed corresponding to testing of unknown objects through the last *n* diffusers used in the final epoch of the training. Furthermore, we observe that the trained diffractive models treat all the diffusers used in the previous epochs (1-99) on average the same (dashed lines in Fig. 5(a)), while the diffusers used in the last epoch (100) are still part of the "memory" of the network as it shows better all-optical reconstruction of unknown test objects through any one of the last *n* diffusers used in the training. Interestingly, due to the small learning rate used at the end of the training phase ($\sim 3\times 10^{-4}$, see the Methods section for details), the diffractive network trained with $n=1$ maintained a fading memory of the last 10 known diffusers. However, this memory did not provide an additional benefit for generalizing to new, unknown diffusers.

Another important observation is that the all-optical reconstruction performance of these trained networks to image unknown test objects through new diffusers is on par with the reconstruction of test objects seen through the diffusers used in epochs 1 through 99 (see Figure 5(b)). These results, along with Figures 2-4, clearly show that these trained diffractive networks have successfully generalized to reconstruct unknown test objects through new random diffusers, never seen before. Figure 5(b) further illustrates that the training strategies that used $n=10$, $n=15$ and $n=20$ perform very similar to each other and are significantly superior to using $n=1$ during the training, as the latter yields relatively inferior generalization and poorer all-optical reconstruction results for unknown new diffusers, as also confirmed in Figs. 6(a-b).

To shed more light on the operation principles of our designed diffractive networks, we also tested the same networks to image distortion-free objects, and therefore removed the random phase diffuser in Fig. 1(a) while keeping all the other components at their corresponding locations; see Figs. 6a and 6b for the resulting images and the PCC values corresponding to the same networks trained with $n=1$, $n=10$, $n=15$ and $n=20$. The fourth column in Fig. 6(a) visually illustrates the diffracted images formed at the output field-of-view of each network, without a diffuser present, demonstrating that the networks indeed converged to a general purpose imager, as also further confirmed by the increased PCC values reported in Fig. 6(b) for the cases without a diffuser.

It is also worth noting that, the diffractive network trained with $n=1$ diffuser per epoch had an easier time to overfit to the last diffuser used during the training phase, and therefore it scored higher when imaging through this last known diffuser (Fig. 6(b)). This is a result of overfitting, which is also evident from its poorer generalization performance under new diffusers as compared to the training strategies that used $n=10$, $n=15$ and $n=20$ diffusers per epoch (see Fig. 6(b)).

**Deeper diffractive networks improve all-optical image reconstruction fidelity**

We also analyzed the impact of deeper diffractive networks that are composed of a larger number of trainable diffractive surfaces on their all-optical reconstruction and generalization performance to see through diffusers. Figure 7 compares the average PCC values for the all-optical reconstruction of unknown test objects using diffractive networks that are designed with different number of diffractive layers. Our results reveal that, with additional trainable diffractive layers, the average PCC values calculated with test images distorted by both known and new random diffusers increase, demonstrating a depth advantage for all-optical image reconstruction.

## Discussion

As demonstrated in our numerical and experimental results, a diffractive network trained with MNIST dataset can all-optically reconstruct unknown resolution test targets through new random diffusers, both of which were not included in the training dataset; these results confirm that the trained diffractive networks do not perform dataset-specific reconstruction, but serve as a general-purpose imager that can reconstruct objects through unknown diffusers. The same conclusion is further supported by the fact that once the diffuser is eliminated from the same set-up, the trained diffractive networks still provide a correct image of the sample at their output, in fact with improved reconstruction fidelity (see Fig. 6).

To further demonstrate the generalization of the all-optical image reconstructions achieved by trained diffractive networks, Supplementary Fig. S3 reports the reconstruction of unknown test objects that were seen through a new diffuser, which had a smaller correlation length (~5$\lambda$) compared to the training diffusers (~10$\lambda$); stated differently, not only the randomly generated test diffuser was not used as part of the training, but also it included much finer phase distortions compared to the diffusers used in the training. The results presented in Supplementary Fig. S3 reveal that, despite a reduction in image contrast, the test objects can still be faithfully reconstructed at the output of the same diffractive network designs using a new diffuser with a smaller correlation length, further deviating from the training phase.

All the results presented in this paper are based on optically thin phase diffusers, which is a standard assumption commonly used in various related studies (*4*, *11*, *75–77*). As a result of this assumption, our results ignore multiple scattering within a volumetric diffuser. Future work will include training diffractive networks that can generalize over volumetric diffusers that distort both the phase and amplitude profiles of the scattered fields at the input plane of a diffractive network. In reality, our experiments already include 3D-printed diffusers that present both phase and amplitude distortions due to the absorption of the THz beam as it passes through different thicknesses of individual features of a fabricated diffuser. Considering the fact that the training of the diffractive networks only included random phase diffusers, the success of our experimental results with 3D-printed diffusers indicate the robustness of this framework to more complex diffuser structures not included in the training. Extensions of this work to all-optically reconstruct object information passing through volumetric diffusers might form the basis of a new generation of imaging systems that can see through e.g., tissue scattering, clouds, fog, etc. at the speed of light, without the need for any digital computation. Hybrid systems that utilize diffractive networks as a front-end of a jointly-trained electronic neural network (back-end) (*66, 68*) is another exciting future research direction that will make use of the presented framework to see through more complicated, dynamic scatters. Finally, our results and presented method can be extended to other parts of the electromagnetic spectrum including e.g., visible/infrared wavelengths, and will open up various new applications in biomedical imaging, astronomy, astrophysics, atmospheric sciences, security, robotics, and many others.

## Methods

**Terahertz continuous wave scanning system**

The schematic diagram of the experimental setup is given in Fig. 1(e). Incident wave was generated through a WR2.2 modular amplifier/multiplier chain (AMC), and output pattern was detected with a Mixer/AMC, both from Virginia Diode Inc. (VDI). A 10 dBm sinusoidal signal at 11.111 GHz (fRF1) was sent to the source as RF input signal and multiplied 36 times to generate continuous-wave (CW) radiation at 0.4THz, and another 10 dBm sinusoidal signal at 11.083 GHz (fRF2) was sent to the detector as a local oscillator for mixing, so that the down-converted signal was at 1 GHz. A horn antenna compatible with WR 2.2 modular AMC was used. We electrically modulated the source with a 1 kHz square wave. The source was put far enough from the input object so that the incident beam can be approximated as a plane wave. A customized reflector is added to the horn antenna to further suppress the reflection noise. The resulting diffraction pattern at the output plane of the network was scanned by a single-pixel detector placed on an XY positioning stage. This stage was built by placing two linear motorized stages (Thorlabs NRT100) vertically to allow precise control of the position of the detector. The output IF signal of the detector was sent to two low-noise amplifiers (Mini-Circuits ZRL-1150-LN+) to amplify the signal by 80 dBm and a 1 GHz (+/-10 MHz) bandpass filter (KL Electronics 3C40-1000/T10-O/O) to get rid of the noise coming from unwanted frequency bands. The amplified signal passed through a tunable attenuator (HP 8495B) and a low-noise power detector (Mini-Circuits ZX47-60), then the output voltage was read by a lock-in amplifier (Stanford Research SR830). The modulation signal was used as the reference signal for the lock-in amplifier. We performed calibration for each measurement by tuning the attenuation and recorded the lock-in amplifier readings. The raw data were converted to linear scale according to the calibration.

**Random diffuser design**

A random diffuser is modeled as a pure phase mask, whose transmittance $t_D(x,y)$ is defined using the refractive index difference between air and diffuser material ($\Delta n \approx 0.74$) and a random height map $D(x,y)$ at the diffuser plane, i.e.,

$$t_D(x,y) = \exp\left(j\frac{2\pi\Delta n}{\lambda}D(x,y)\right) \quad (2)$$

where $j = \sqrt{-1}$ and $\lambda = 0.75$mm. The random height map $D(x,y)$ is further defined as (*11*)

$$D(x,y) = W(x,y) * K(\sigma). \quad (3)$$

where $W(x,y)$ follows normal distribution with a mean $\mu$ and a standard deviation $\sigma_0$, i.e.

$$W(x,y) \sim \mathcal{N}(\mu, \sigma_0). \tag{4}$$

$K(\sigma)$ is the zero mean Gaussian smoothing kernel with standard deviation of $\sigma$. '$*$' denotes the 2D convolution operation. In this work, we chose $\mu = 25\lambda$, $\sigma_0 = 8\lambda$ and $\sigma = 4\lambda$ to randomly generate the training and testing diffusers, mimicking glass-based diffusers used in the visible part of the spectrum. For this choice of diffuser parameters, we further calculated the mean correlation length ($L$) using a phase-autocorrelation function $R_d(x,y)$ that is defined as (78)

$$R_d(x,y) = \exp(-\pi(x^2 + y^2)/L^2). \tag{5}$$

Based on 2000 randomly generated diffusers with the above described parameters and their corresponding phase-autocorrelation functions, we determined the average correlation length as $\sim 10\lambda$. Different from these diffusers, for Supplementary Fig. S3, we used $\sigma = 2\lambda$ to randomly generate phase diffusers with an average correlation length of $L = \sim 5\lambda$.

The difference between two randomly-generated diffusers are quantified by the average pixel-wise absolute phase difference, i.e., $\Delta\phi = \overline{|(\phi_1 - \overline{\phi_1}) - (\phi_2 - \overline{\phi_2})|}$, where $\phi_1$ and $\phi_2$ represent the 2D phase distributions of two diffusers, and $\overline{\phi_1}$ and $\overline{\phi_2}$ are the mean phase values of each. When we randomly generate new phase diffusers, it can be regarded as a novel/unique diffuser when $\Delta\phi > \pi/2$ compared to all the existing diffusers randomly created before that point.

**Forward propagation model**
A random phase diffuser defined in Eq. (2) positioned at $z_0$ provides a phase distortion: $t_D(x,y)$. Assuming that a plane wave is incident at an amplitude-encoded image $h(x,y,z=0)$ positioned at $z = 0$, we modeled the disturbed image as:

$$u_0(x,y,z_0) = t_D(x,y) \cdot [h(x,y,0) * w(x,y,z_0)] \tag{6}$$

where,

$$w(x,y,z) = \frac{z}{r^2}\left(\frac{1}{2\pi r} + \frac{1}{j\lambda}\right)\exp\left(\frac{j2\pi r}{\lambda}\right) \quad (7)$$

is the propagation kernel following the Rayleigh-Sommerfeld equation (*63*) with $r = \sqrt{x^2 + y^2 + z^2}$. The distorted image is further used as the input field to the subsequent diffractive system. The transmission of layer $m$ (located at $z = z_m$) of a diffractive system provides a field transmittance:

$$t_m = \exp(j\phi(x,y,z_m)). \quad (8)$$

Being modulated by each layer, the optical field $u_m(x,y,z_m)$ right after the $m^{\text{th}}$ diffractive layer positioned at $z = z_m$ can be formulated as

$$u_m(x,y,z_m) = t_m(x,y,z_m) \cdot [u_{m-1}(x,y,z_{m-1}) * w(x,y,\Delta z_m)] \quad (9)$$

where $\Delta z_m$ is the axial distance between two successive diffractive layers, which was selected as $2.7\lambda$ in this paper. After being modulated by all the $M$ diffractive layers, the light field is further propagated by an axial distance of $\Delta z_d = 9.3\lambda$ onto the output plane, and its intensity is calculated as the output of the network, i.e.,

$$o(x,y) = |u_M * w(x,y,\Delta z_d)|^2. \quad (10)$$

**Network training**

The diffractive neural networks used in this work were designed for λ≈0.75 mm coherent illumination and contain 240×240 pixels on each layer providing phase-only modulation on the incident light field, with a pixel size (pitch) of 0.6 mm. During the training, each hand-written digit of the MNIST training dataset is first upscaled from 28×28 pixels to 160×160 pixels using bilinear interpolation, then padded with zeros to cover 240×240 pixels. $B=4$ different randomly selected MNIST images form a training batch. Each input object $h_b(x,y)$ in a batch is numerically duplicated *n* times and individually disturbed by a set of *n* randomly selected diffusers. These distorted fields are separately forward propagated through the diffractive network. At the output plane, we get *n* different intensity patterns: $o_{b1}, o_{b2} ... o_{bn}$. All $B \times n$ output patterns are collected to calculate the loss function:

$$Loss = \frac{\sum_{b,i=1}^{b=4,i=n}\{-P(o_{bi}, h_b) + E(o_{bi}, h_b)\}}{B \times n} \qquad (11)$$

In Eq. (10) $P(o_{bi}, h_b)$ denotes the PCC between the output and its ground truth image $h_b$, calculated based on Eq. (1). Furthermore, $E(o_{bi}, h_b)$ denotes an object-specific energy efficiency-related penalty term, defined as:

$$E(o_{bi}, h_b) = \frac{\sum_{x,y}(\alpha(1-\widehat{h_b}) \cdot o_{bi} - \beta \widehat{h_b} \cdot o_{bi})}{\sum_{x,y} \widehat{h_b}}. \qquad (12)$$

In Eq. (11) $\widehat{h_b}$ is a binary mask indicating the transmittance area on the input object, defined as:

$$\widehat{h_b}(x,y) = \begin{cases} 1, & h_b(x,y) > 0 \\ 0, & otherwise \end{cases}, \qquad (13)$$

where $\alpha$ and $\beta$ are hyper-parameters that are optimized to be 1 and 0.5 respectively.

The resulting loss value (error) is then back-propagated and the pixel phase modulation values are updated using the Adam optimizer (*79*) with a decaying learning rate of $Lr = 0.99^{Ite} \times 10^{-3}$, where $Ite$ denotes the current iteration number. Our models were trained using Python (v3.7.3) and TensorFlow (v1.13.0, Google Inc.) for 100 epochs with a GeForce GTX 1080 Ti graphical processing unit (GPU, Nvidia Inc.), an Intel® Core ™ i9-7900X central processing unit (CPU, Intel Inc.) and 64 GB of RAM. Training of a typical diffractive network model takes ~24 hours to complete with 100 epochs and *n*=20 diffusers per epoch. The phase profile of each diffractive layer was then converted into the height map and corresponding .stl file was generated using MATLAB, and subsequently 3D printed using Form 3 3D printer (Formlabs Inc., MA, USA).

**Quantification of the reconstructed resolution test target period**

For an amplitude-encoded, binary resolution test target (with a period of $p$) the transmission function can be written as:

$$h_t(x,y) = \begin{cases} 1, & x \in \left(-\frac{5}{2}p, -\frac{3}{2}p\right) \cup \left(-\frac{p}{2}, \frac{p}{2}\right) \cup \left(\frac{3}{2}p, \frac{5}{2}p\right) \\ 0, & otherwise \end{cases} \quad (14)$$

The diffractive network forms the reconstructed image $o(x,y)$ of the resolution test target at the output field-of-view, over an area of X×Y mm². To quantify/measure the period of the reconstructed test targets, the intensity was first averaged along the y axis, yielding a 1D intensity profile:

$$l(x) = \frac{\int_{y=0}^{Y} o(x,y) \cdot dy}{Y}. \quad (15)$$

Subsequently we fit a curve $F(x)$ to $l(x)$ by solving:

$$\underset{a_1,a_2,a_3,b_1,b_2,b_3,c_1,c_2,c_3}{\operatorname{argmin}} \left(\sum ||F(x) - l(x)||^2\right), \quad (16)$$

where

$$F(x) = a_1 \exp\left(-\left(\frac{x-b_1}{c_1}\right)^2\right) + a_2 \exp\left(-\left(\frac{x-b_2}{c_2}\right)^2\right) + a_3 \exp\left(-\left(\frac{x-b_3}{c_3}\right)^2\right). \quad (17)$$

The measured/resolved period ($\hat{p}$) at the output image plane is then calculated as:

$$\hat{p} = \frac{\max(b_1,b_2,b_3) - \min(b_1,b_2,b_3)}{2}. \quad (18)$$

**Image contrast enhancement**

For the purpose of better image visualization, we digitally enhanced the contrast of each experimental measurement using a built-in MATLAB function (*imadjust*), which by-default saturates the top 1% and the bottom 1% of the pixel values and maps the resulting image to a dynamic range between 0 and 1. The same default image enhancement is also applied to the results shown in Figs. 1(b-c), 4 and Supplementary Figs. S2 and S3. All quantitative data analyses, including PCC calculations and resolution test target period quantification results, are based on raw data, i.e., did not utilize image contrast enhancement.

**Lens-based imaging system simulation**

We numerically implemented a lens-based imaging system to evaluate the impact of a given random diffuser on the output image; see e.g., Fig. 1(b) and Supplementary Fig. S2. A Fresnel lens was designed to have a focal length ($f$) of 145.6 $\lambda$ and a pupil diameter of 104 $\lambda$ (*80*). The transmission profile of the lens $t_L$ was formulated as:

$$t_L(\Delta x, \Delta y) = A(\Delta x, \Delta y) \exp\left(-j\frac{\pi}{\lambda f}(\Delta x^2 + \Delta y^2)\right), \qquad (19)$$

where $\Delta x$ and $\Delta y$ denote the distance from the center of the lens in lateral coordinates. $A(\Delta x, \Delta y)$ is the pupil function, i.e.,

$$A(\Delta x, \Delta y) = \begin{cases} 1, & \sqrt{\Delta x^2 + \Delta y^2} < 52\lambda \\ 0, & otherwise \end{cases}, \qquad (20)$$

The lens was placed $2f$ (291.2$\lambda$) away from the input object. The input object light was first propagated axially for $z_0 = 53\lambda$ to the random diffuser plane using the angular spectrum method. The distorted field through the diffuser was then propagated to the lens plane, and after passing through the lens the resulting complex field was propagated to the image plane ($2f$ behind the lens), also using the angular spectrum method. The intensity profile at the image plane was calculated as the resulting image, seen through an aberration-free lens, distorted by a random phase diffuser.

# Figures

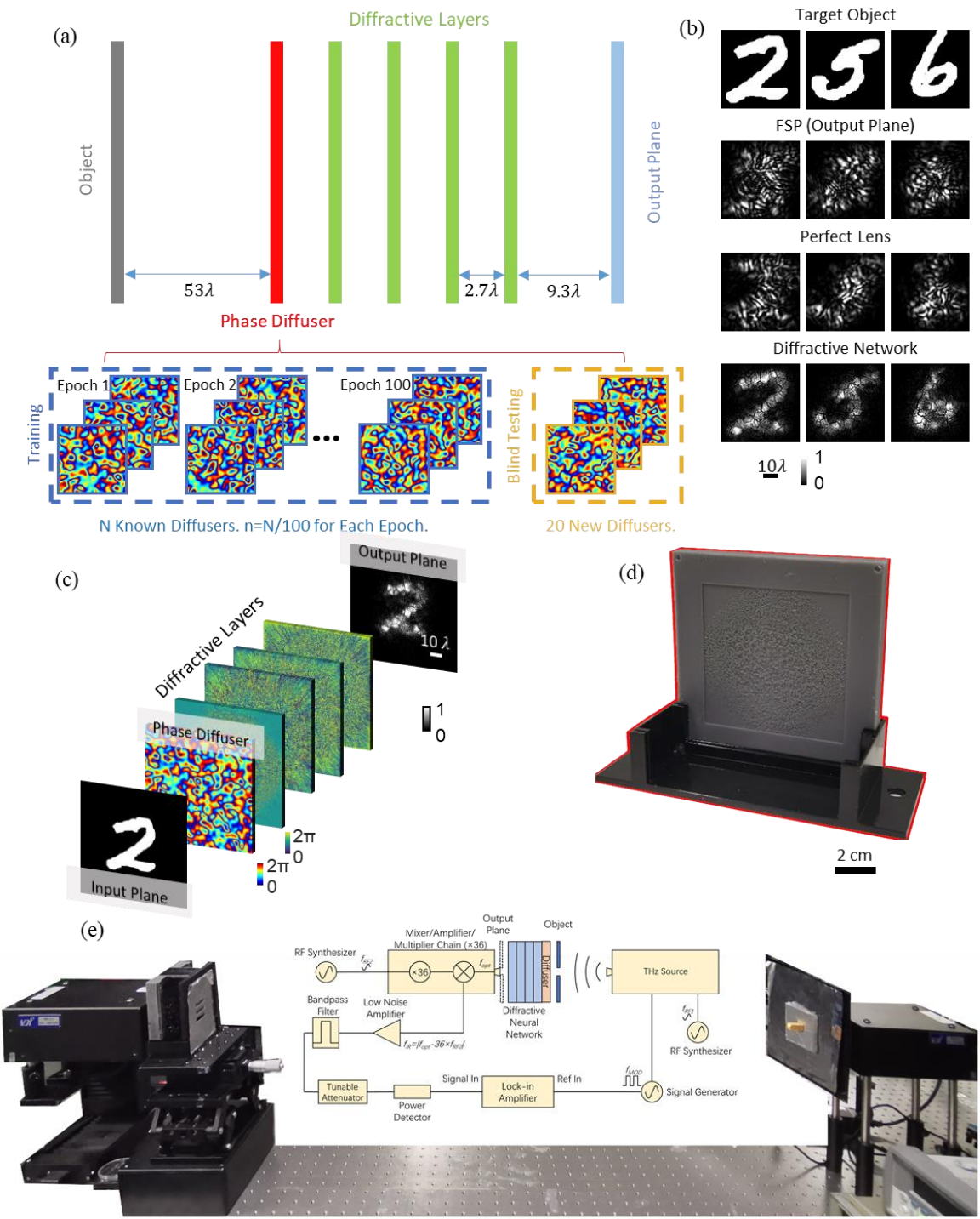

**Figure 1. All-optical imaging through diffusers using diffractive surfaces. a**. Training and design schematic of a 4-layered diffractive system that can see through unknown/new randomly generated phase diffusers. **b**. Sample images showing the image distortion generated by random

diffusers. Top: input images. Second row: free-space propagation (FSP) of the input objects through the diffuser, without the diffractive layers, imaged at the output plane. Third row: the input objects imaged by an aberration-free lens through the diffuser. Fourth row: the outputs of the trained diffractive network. **c**. Schematic of a 4-layered network trained to all-optically reconstruct the input field of view seen through an unknown random diffuser. **d**. The photograph of the 3D printed network shown in (c). **e**. Schematic and photograph of the experimental apparatus used for testing the design shown in (c) using continuous wave coherent THz illumination.

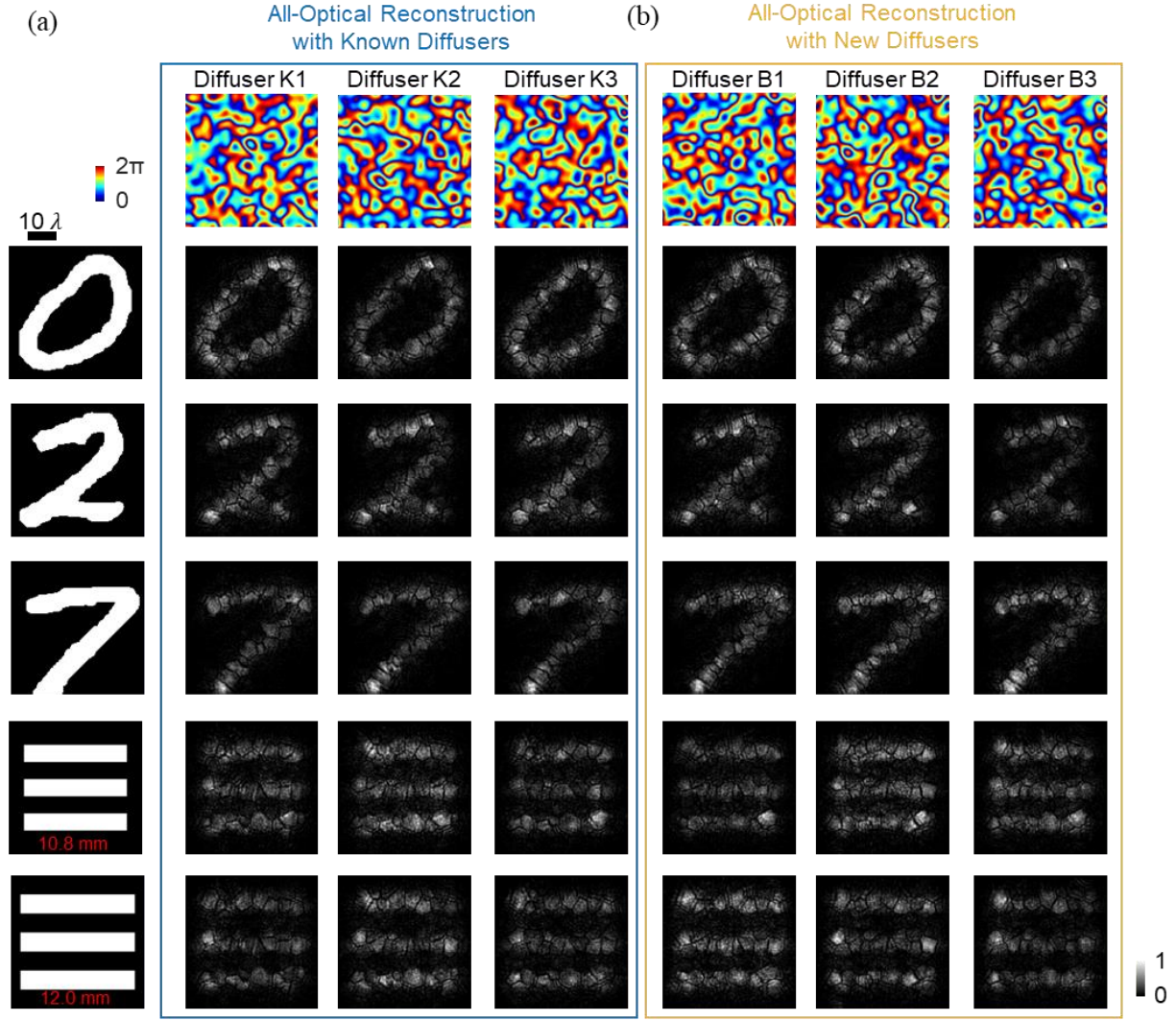

**Figure 2. Simulation results of the all-optically reconstructed images of test objects distorted by (a) known and (b) new diffusers using the trained diffractive network shown in Fig. 1(c)**.

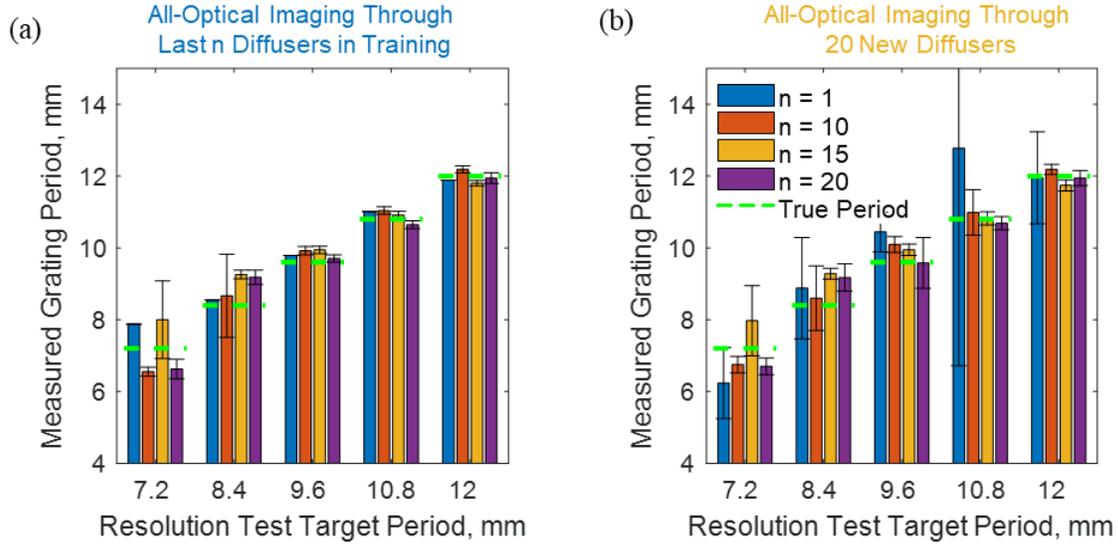

**Figure 3. Generalization of diffractive networks that were trained with MNIST image data to reconstruct the images of different resolution test targets, seen through (a) known and (b) new randomly generated diffusers.** Despite the fact that such resolution test targets or similar line-based objects were never seen by the networks during their training, their periods are successfully resolved at the output plane of the diffractive networks.

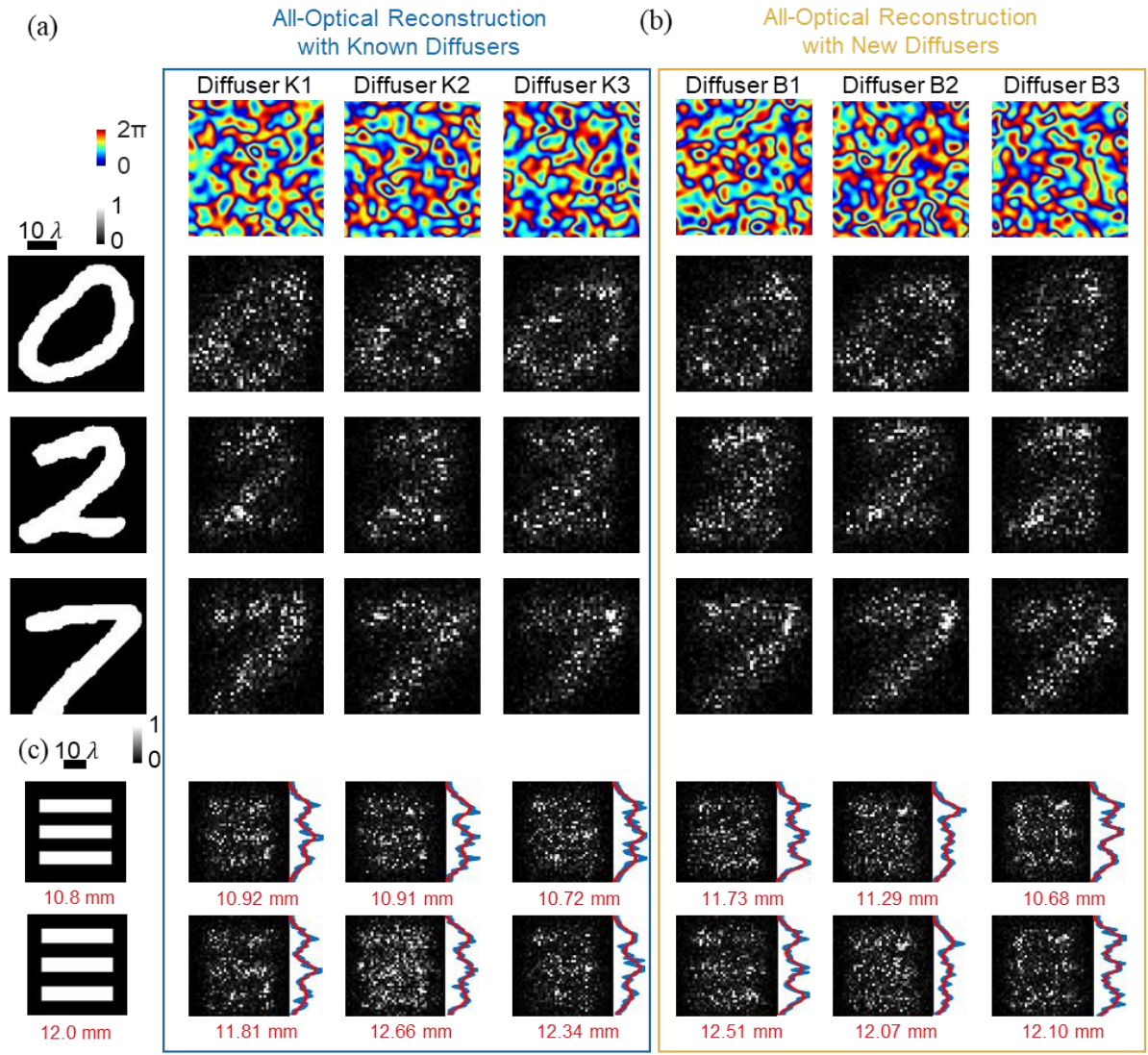

**Figure 4.** Experimental results of the all-optically reconstructed images of test objects distorted by (a) known and (b) new diffusers using the trained diffractive network shown in Fig. 1(c). (c) The measured periods of the resolution test targets imaged through known and unknown diffusers are labeled in red.

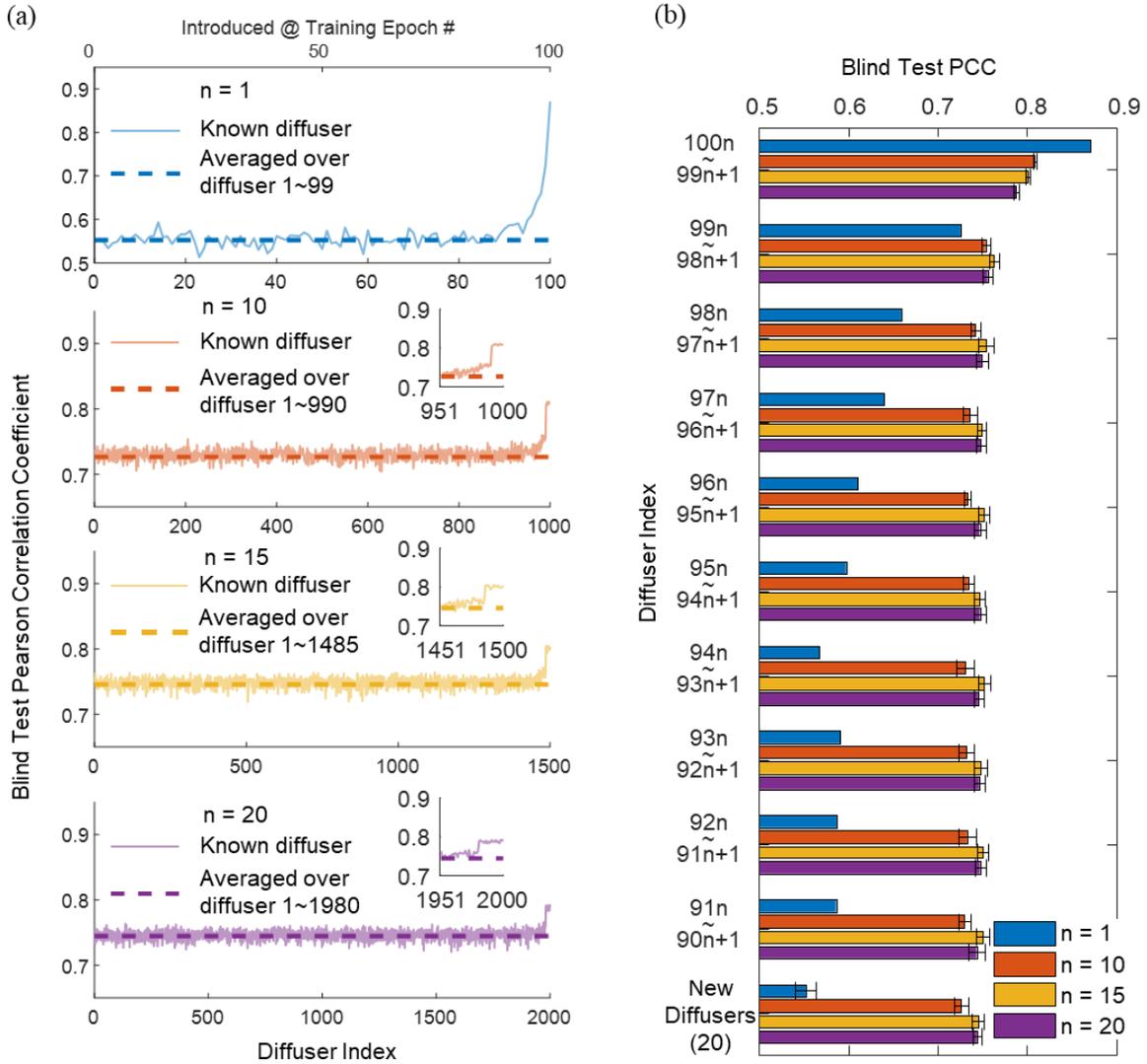

**Figure 5. Memory of diffractive networks. a**. After being trained for 100 epochs, all the finalized networks ($n$ = 1, 10, 15 and 20) were compared by calculating the average PCC values over unknown objects distorted by *all* the known diffusers (solid line). Dashed line: the average PCC value over unknown objects distorted by diffusers indexed as 1-99$n$. Inserts: the same plot zoomed into the last 50 diffusers. **b**. PCC values of each finalized network tested with images distorted by 10$n$ diffusers used in last 10 epochs in training ($n$=1, 10, 15 and 20, respectively) and 20 new random diffusers (never seen before). The error bars reflect the standard deviation over different diffusers.

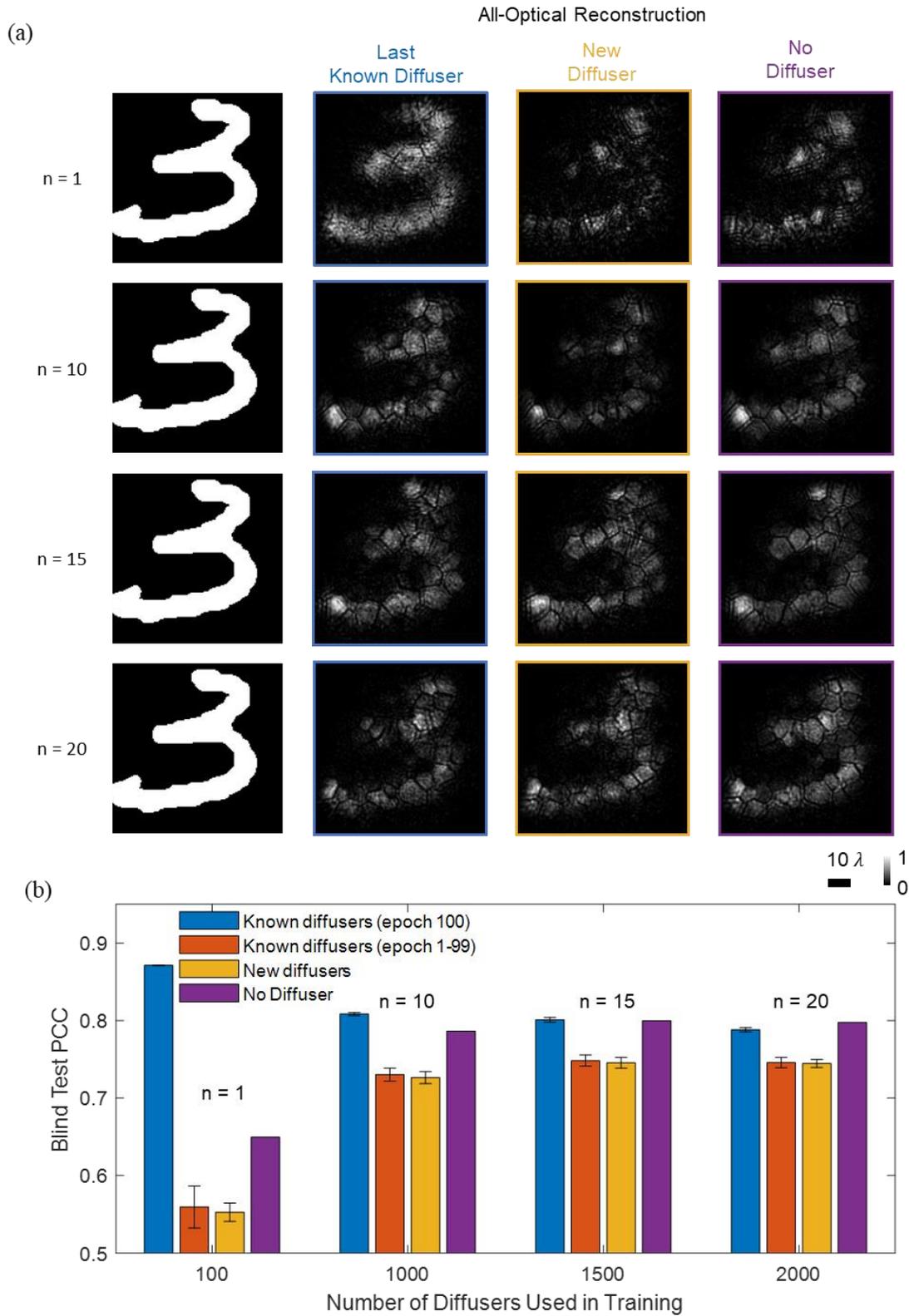

**Figure 6. Comparison of diffractive network output images under different conditions. a.** Output images corresponding to the same input test object imaged through diffractive networks

trained with $n=1$, $n=10$, $n=15$ and $n=20$. Second column: imaged through a known diffuser; third column: imaged through a new diffuser; fourth column: imaged without a diffuser. **b.** The PCC values corresponding to the networks trained with $n = 1, 10, 15$ and $20$ over input test objects distorted by known diffusers, new diffusers, as well as imaged without a diffuser. The error bars reflect the standard deviation over different diffusers.

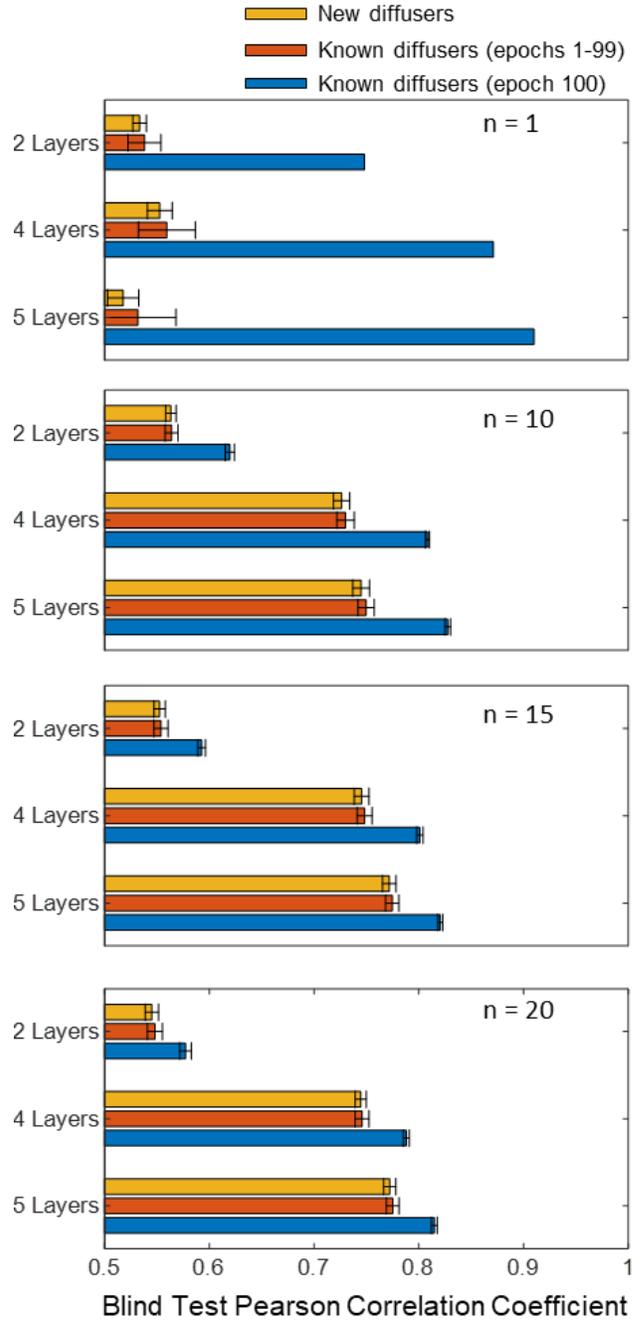

**Figure 7. Additional trainable diffractive surfaces improve the all-optical image reconstruction of objects seen through unknown random diffusers**. The error bars reflect the standard deviation over different diffusers.